\documentclass[fleqn,12pt]{SelfArx} 

\definecolor{color1}{RGB}{0,0,185} 
\definecolor{color2}{RGB}{0,20,20} 

\usepackage{hyperref} 
\hypersetup{hidelinks,colorlinks,breaklinks=true,urlcolor=color2,citecolor=color1,linkcolor=color1,bookmarksopen=false,pdftitle={Title},pdfauthor={Author}}

\JournalInfo{Accepted to Solar Physics, 2015} 

\PaperTitle{
The Origin of Superflares on G-Type Dwarf Stars of Various Ages
} 

\Authors{M. M. Katsova\textsuperscript{1}*, M. A. Livshits\textsuperscript{2}} 
\affiliation{\textsuperscript{1}\textit{
Sternberg Astronomical Institute,
Lomonosov Moscow State University,
Moscow, Russia}} 
\affiliation{\textsuperscript{2}\textit{
Pushkov Institute of Terrestrial Magnetism, Ionosphere, and Radio-Wave Propagation,\\
 ~~ Russian Academy of Sciences, Moscow, Troitsk, Russia
}} 
\affiliation{* \textbf{Corresponding author}: maria@sai.msu.ru} 

\newcommand{\keywordname}{Keywords} 

\Abstract{
We analyze new observations of superflares on G stars discovered in the optical and near IR-ranges with the Kepler mission. An evolution of solar-type activity is discussed. We give an estimate of the maximal total energy, $E_{tot} = 10^{34}$\;erg of a flare that can occur on the young Sun at its age of 1~Gyr when the cycle was formed. We believe that the main source of the flare optical continuum is a low-temperature condensation forming in the course of the response of the chromosphere to an impulsive heating. For a superflare on the young Sun, we adopt the accelerated electron flux, 
$F_e (E>\mbox{20 keV}) = 3 \times 10^{11} \: \mbox{erg} \; \mbox{cm}^{-2} \; \mbox{s}^{-1}$, 
that is limited by the return current, and obtain the area of the optical continuum source on a G star, 
$S \approx 10^{19} \:\mbox{cm}^2$. 
This value is close to the area of the $H_\alpha$-ribbons in the largest solar flares, while the area of bright patches of a white-light flare on the contemporary Sun is smaller by about two orders of magnitude. At the same electron flux and the hard electron spectrum, the stellar flare of the similar energy should be accompanied by the microwave source of about 2~mJy at frequencies 10--100~GHz at a distance of 100~pc. We discuss the possible detection of the flare-produced lithium in the course of spallation reactions. The detection of the flare microwave source and the emission in the Li resonant line could demonstrate how effective can be particle acceleration on stars in the lower part of the main sequence. 
}

\begin{document}

\flushbottom 

\maketitle 

\tableofcontents 

\thispagestyle{empty} 


\section{Introduction} 


The first observations of powerful flares of the optical continuum in G stars give evidence for acceleration of huge amount of particles, in particular, electrons. These results of the Kepler mission were presented by Maehara et al. (2012) who reported on total bolometric energies of the flare emission from $10^{33}$ to $10^{36}$\:erg. 

Maehara et al. (2012) presented observations of 148 G stars which produced 365 superflares with a total energy $> 3 \times 10^{33}$\:erg registered during 120 days. The Kepler mission monitoring was done in the V and R bands and in two time-resolution regimes, short-cadence, 1 min, and long-cadence, 30 min, and the first results were based on a set of data with the lower time resolution.

The energies of white-light stellar flares were re-calculated in the bolometric values in black-body approximation. Maehara et al. (2012) estimated the frequency of occurrence of flares with energies from $10^{33}$ to $10^{36}$\:erg, to be somewhat higher for the cooler G stars ($T_{\it eff} = 5100-5600$\:K) than that for the hotter G stars ($T_{\it eff} =  5600-6000$\:K). Superflares were found to occur more often on G stars rotating with periods from a few hours to several days; a frequency of these events on stars with rotational periods $>10$\;days is several times smaller. 

The further analysis of the Kepler data of monitoring during 500 days carried out by Shibayama et al. (2013) and Balona (2015). The results by Shibayama et al. (2013) based on a longer series of observations confirm those obtained by Maehara et al. (2012), give a reliable flare frequency, and provide evidence that superflares occur quite rarely on G stars. Note that Balona (2015) found that flare energies are strongly correlated with stellar luminosity and radius. 

Detection of such powerful non-stationary processes on normal late-type stars raises a question about the principal possibility of such phenomena that may require extreme physical conditions like strong magnetic fields, close companions (a star or a planet), comet impact etc. But, here we do not go beyond a standard knowledge about the physics of flares. The most effective particle acceleration takes place during some of the largest flares on the Sun. These events are known well and are studied in detail. So, by analyzing the white-light solar flare on 14 July 2000, Livshits and Belov (2004) assumed that the source of this effective acceleration is situated in close proximity to a sunspot. This is confirmed by the analysis of other powerful events like those in August 1972 (Svestka, 1976) and 28 October 2003; but most clearly, it was seen during the flare on 20 January 2005 above sunspot umbrae (Grechnev et al., 2008). These high-energy phenomena are widely under discussion (see, for instance, the review by Ramaty, 1986; Fletcher et al., 2011). Before the Kepler observations, it was hard to imagine that the acceleration in moderate magnetic fields, observed on normal G stars, can be much more efficient than in large solar flares. 

We suppose that the basic property of superflares is an efficient acceleration of electrons and protons. These particles are a cause of both line emission and optical continuum in solar flares. In superflares, this continuum emission is extremely intense. The dynamic response of the solar chromosphere to impact by accelerated electrons leads to the formation of the hard X-ray radiation, line and continuum emission, in particular, in optical wavelengths. The explosive evaporation of the chromosphere occurs in many large impulsive flares on the Sun and red dwarf stars. We believe that this process takes place also in superflares.

           The aim of this work is to understand such strong optical continuum emission in the frameworks of the current knowledge about flares on the Sun and red dwarfs. 

          The solution of this basic problem requires information about G stars with superflares.  As a rule, these stars are younger than the present Sun. The current experience of flare studies relates to phenomena observed on the Sun and other G stars of the comparable age. Therefore, one needs to make an educated guess where superflares occur and how they are associated with other active phenomena on these stars. 

When the Sun was younger than 1 Gyr, its non-stationary processes could differ significantly from the solar events in the contemporary epoch. The general character of activity of late-type stars depends on their age. The activity level of open cluster members is noticeably higher than in the present Sun. The problem of the evolution of stellar activity is widely discussed after the article by Barnes (2003).  A method for the estimation of the stellar age from their activity level (a single-parametric gyrochronology) was suggested by Mamajek and Hillenbrandt (2008). For a set of phenomena as those observed on the Sun today, including formation of the regular cycle, we use the term ``solar-type activity''. 

One can expect that such kind of activity was formed on a main-sequence G star to the age slightly older than 1 Gyr (see, for example, Katsova and Livshits, 2011; 2014). On young stars of ages from 50 to 500 Myr, the character of the magnetic fields and non-stationary processes differ from the solar ones. The basic distinction is due to a greater impact of the large-scale magnetic fields on these objects, in particular, on the longer duration of powerful flares. In this context, the Kepler observations of superflares can shed light on age-related changes of flare activity from young T Tau-type stars to the young Sun. 

So far observations have demonstrated that characteristics of these events, besides their total energy, do not differ significantly from the largest solar analogues. Indeed, their duration in the optical is from a few minutes to one or several hours, the temporal profiles are similar to many X-class flares on the Sun. This was a reason why the first theoretical analysis by Shibayama et al. (2013) and some other works used simple scaling relationships. Some disadvantage of this approach is that for explanation of the observed powerful phenomena, one needs to assume that the area occupied by spots, is close to 100\% of the stellar surface. Note that the relation between spot areas and the flare energy was investigated recently by Aulanier et al. (2013). Of course, because observed values of stellar spots on young G star do not exceed 30\% (Berdyugina, 2005), the simple scaling relationship by Shibayama et al. (2013) cannot be applied to explain superflares of the largest energies. It is necessary here to analyze possibilities of generation of strong optical emission that should be more effective as compared to those on the contemporary Sun. 

We consider a few outstanding problems. The first of them relates to the question whether it is possible to explain the strong optical continuum observed during superflares in the framework of the gas-dynamic response of the chromosphere to an impact of accelerated particles. The second: are superflares associated with the efficient acceleration of particles? This can be verified through their manifestations in flares. We evaluate the microwave radiation of superflares and discuss the possibility of its detection. Then, we discuss under what conditions nuclear spallation reaction can occur during superflares and estimate the resulting expected enhancement of the resonant Li lines at 6708\,\AA.

Besides, the question posed in the title of the article by Shibata et al. (2013): ``Can superflares occur on our Sun'' is of interest. The answer requires an analysis of the relation between development of flares and the magnetic fields, which have recently been observed on G stars recently (Marsden et al. 2014). We formulate the problem as follows. Are superflares similar to solar phenomena? Are there hypothetical possibilities for explanation of events with such power, which cannot in principle occur on the Sun, while they are observed on the youngest stars?

\section{Evaluation of the accelerated electron fluxes  for flares registered with the Kepler mission}

First of all, let us discuss the problem of formation of the strong flare optical continuum. Previously it was thought that white-light flares occur very rarely on the Sun. However, the Yohkoh observations (Matthews et al., 2003) and the data obtained with TRACE (Hudson, Wolfson, and Metcalf, 2006; Fletcher et al., 2007) in the blue spectral range have demonstrated a greater prevalence of these phenomena.

Besides, a great progress has been achieved in the last years in the study of the X-ray radiation of flares. One should mention particularly the results of the RHESSI observations, which permitted to obtain comprehensive information about the accelerated electron fluxes. These particles can be responsible for heating and the radiation of the flares in different spectral ranges. The sudden enhancement of the optical continuum has also been observed during flares on red dwarf stars (Gershberg, 2005). Modern observations of stellar flares give evidence for both the common properties and some differences of physical processes on the Sun and red dwarfs (Schmitt et al., 2008; Kowalski et al., 2013).

The Kepler observations of late-type stars were performed in the optical range 4300--8300\,\AA. We will use the total energies of optical flare emission presented by Shibayama et al. (2013). We assume that the basic source of the optical continuum of large solar flares is a low-temperature condensation, which is formed as a result of the gas-dynamical response of the chromosphere to the impulsive heating by the accelerated electron flux. As known, the first numerical modeling of this process was carried out by Kostyuk and Pikelner (1975). This approach was further developed by Livshits et al. (1981) in the article entitled ``The optical continuum of solar and stellar flares''. The corresponding model of impulsive flares on red dwarf stars was developed by Katsova, Boiko, and Livshits (1997). An exhaustive physical justification of individual aspects of the process of explosive evaporation was given by Fisher, Canfield, and McClymont (1985). We recall that the gas-dynamical response includes the primary heating of a small area of the chromosphere, formation of two shock waves propagating upward and downward, and the formation of the low-temperature condensation between the shock front and downward-moving temperature jump. This is accompanied by an upward outflow of hot plasma, and as a result hot gas gradually fills the coronal loops. Over the past years there have appeared a lot of articles with interpretation of individual events on the Sun and the stars, and now we can assume that this process is actually implemented in impulsive flares (see, for instance, Antonucci, 1989; Liu et al., 2006; Martinez Oliveros et al., 2012).

When modeling processes of a superflare with an extremely high total energy, one has to take into account two main problems. First, at large fluxes of the accelerated electrons, a return current arises, that prevents energy to penetrate into deep chromospheric layers. The gas-dynamical modeling with return current was performed by Karlicky and Henoux (1992). This complicated problem was further considered by Matthews, Brown, and van Driel-Gesztelyi (1998); Zharkova (2000); Lee, B\"uchner, and Elkina (2008). Calculations show that this electron flux is so large possibly due to processes in outer atmosphere of the Sun (Boiko and Livshits, 1999). In our calculations we assume that the optical continuum emission source arises in powerful events when the energy of directly precipitating electrons Fe  reaches $3\times 10^{11}\,\hbox{erg}\:\hbox{cm}^{-2}\:\hbox{s}^{-1}$. Here, for definiteness, we assume that the low-energy cutoff of these electrons is 20 keV in X-ray spectra of the largest flares.

Second, such strong electron fluxes during impulsive events are a source of perturbations that can penetrate as deeply as the region of the temperature minimum, i.e., to the upper boundary of the photosphere. When we modeled solar impulsive flares, we assumed that the radiative losses were determined mainly by the hydrogen line emission. In superflares, physical conditions in a low-temperature condensation are as follows: the distribution of the hydrogen atoms over the energy levels reaches equilibrium, and the energy losses in the hydrogen lines are not enough to compensate for the heating inside the condensation. Therefore, we have to consider motions of the plasma, the excitation and ionization of the hydrogen in a homogeneous layer behind the downward moving shock front in solar and stellar flares. The optical depth of this condensation in the optical continuum is close to one for the most powerful events.

Here the new advances in the elementary processes in partially ionized plasma are taken into account. The transition from the optically thin to optically thick plasma in the optical continuum of a low-temperature condensation is studied in detail by Morchenko, Bychkov, and Livshits (2015) for the atmosphere of a red dwarf (when the influence of the photosphere radiation can be neglected). We apply the two-temperature approximation where the electrons and ions are heated differently behind the shock front. In our gas-dynamics model, the condensation is presented as a homogeneous planar layer with a fixed thickness of 10\,km. The electron temperature is 11600\,K (1\,eV). At high densities the optical spectrum represents intensified optical continuum, against which the lines are seen as small maxima of their centers. As an example, the results of modeling the optical continuum are given in Figure 1: the lower curve corresponds to a density of $7\times 10^{15}\,\hbox{cm}^{-3}$, and the upper curve is for $3\times 10^{16}\,\hbox{cm}^{-3}$. It is clear that the intensity ratio before and behind the Balmer limit decreases, which means that the Balmer jump is diminished. The optical thickness at $\lambda=4170$\,\AA\ is 1.05 and 5.79 for both cases correspondingly. So, the emission of this 10 km-layer at $T = 11600$\,K and $n=3\times 10^{16}\,\hbox{cm}^{-3}$ almost becomes black-body radiation.

\begin{figure}[!t]\centering                      
\includegraphics[width=0.7\linewidth]{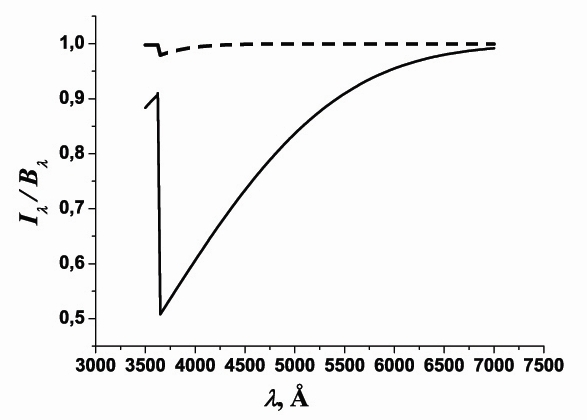}
\caption{\\
The theoretical spectrum of the optical continuum of a homogeneous 10 km-layer of hydrogen plasma with the temperature of 11600~K. The intensity of the optical radiation of this layer to the corresponding black-body radiation ratio, $I_\lambda/B_\lambda(T)$, is given for densities $7\times 10^{15}\:\mbox{cm}^{-3}$ and $3\times 10^{16}\:\mbox{cm}^{-3}$ for the lower and the upper (dashed) lines correspondingly. The optical thickness for both cases are 1.05 and 5.79 at $\lambda = 4170$\,\AA.
}
\end{figure}

This approach permits us to evaluate the contribution of the radiation of the gas-dynamical condensation to the optical continuum of solar and stellar flares of different energies. Observations of superflares allow us to evaluate the area of the optical emission source in the frameworks of the gas-dynamics model. In powerful events, the radiation approaches to a black-body one, and the total emitted energy is

$$
 E_{\it flare} = \sigma T^4 S \Delta t, \eqno (1)
$$

where $S$ is the area of the optical source and $\Delta t$ is the duration of the radiation. 
As we show below, the evolution of the magnetic fields on a G star with the $R=R_\odot$ can lead to flares with a maximal total energy $10^{34}\,$erg. Our point is that we make no distinction between the blue and the red continuum and adopt the temperature of the emission as 10000\,K. Then from the expression (1) for a flare lasting $\Delta t = 0.5\,$hour, we obtain $S = 1.0\times 10^{19}\,\hbox{cm}^2$. This value is slightly higher than the $H_\alpha$-ribbon area of the greatest solar flares. 

It is known that the brightest patches of white-light flares on the Sun occupy an area of about of $2\times 10^{16}\,\hbox{cm}^{-2}$. Now it is confirmed by current data about the area of the brightest footpoints in the G-band of the continuum (Krucker et al. 2011). In large solar flares, the area of the white-light radiation is 100 times smaller than the area of the $H_\alpha$-ribbon emission in the same events. Such a large optical continuum emission area provides evidence that accelerated particles within the loop system continue to maintain this radiation in loop footpoints for some longer time immediately after one or more episodes of an impulsive acceleration. We discuss this in more detail in the next section.

\section{On a model of the optical continuum  source of superflares}

Now we propose a model of the optical continuum source for a superflare with a total energy of 
$10^{34}\,$erg that can occur on the young Sun. This model is based on an analysis of contemporary observations of the largest solar flares. In fact, the study in this direction was started with the Bastille Day flare on 14 July 2000, because it became possible to obtain both EUV-line and X-ray images. So, Livshits and Belov (2004) showed that features of the radiation in different spectral ranges are associated with the extremely efficient particle acceleration occurring at low heights in the immediate vicinity of spot penumbrae. In particular, Livshits and Belov (2004) presented schematically coronal flare loops and ribbons against the photospheric layers of the active region (this is shown in Figure 2). The hard X-ray image and the temporal profile of the coronal emission of the hardest source are also given. Tentatively, three kinds of the hard X-ray sources were revealed. The first one is the region of the main acceleration (I); this is a single or a few very low-lying loops located above the polarity inversion line (PIL) crossing the big spot. The system of coronal loops begins to form above the PIL over the entire active region, and some of footpoints of loops in this post-eruptive arcade are sources of the enhanced X-ray radiation (II). The coronal loop, which connects the region of the main particle acceleration with another ribbon, is often a source of strong microwave radiation, because many particles of relativistic energies are captured therein. The footpoint of this loop (III) is also bright in the hard X-ray and $\gamma$-ranges.

\begin{figure*}[!t]\centering               
\includegraphics[width=\linewidth]{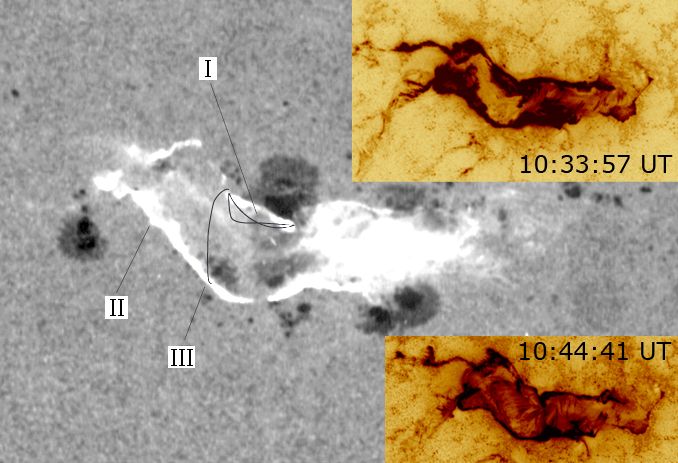}
\caption{\\
A schema of the position of sources I, II, and III of the optical radiation of superflares. Two TRACE images of the flare on 14 July 2000 obtained close to the $\gamma$- burst maximum 10:27 UT, in white light and at 195\,\AA, are superimposed. Two ribbons at 195\,\AA\ emitted weakly at the loop top are seen (the brightness of the western part is reduced). The TRACE images at 1600\,\AA\ are presented for 10:33:57 UT (the upper right corner) and 10:44:41 UT (in the bottom, right corner). The source I is better seen in 10:33:57 UT. The loop connecting sources I and III was gradually stretched to the East during 10 min that can be due to the lifting of its top. The optical continuum of superflares is assumed to form in the flare ribbons and intensifies in sources I, II, and III.
}
\end{figure*}

Note that in previous analysis of data on the Bastille Day flare the X-ray sources I, II, and III were identified as intensive inhomogeneities at 50--200\,keV-images of ribbons (Fletcher and Hudson (2001). These hard X-ray sources can be detected from other observations as well. For the Bastille Day flare it was obtained that the maximum of the radiation of the region I even of the cool coronal plasma ($T\sim 10^6\,$K) at 171\,\AA\   coincided with a peak of the $\gamma$-burst (Aschwanden and Alexander, 2001). This confirms the conclusion that the high-energy events took place in region I, where occurs an effective acceleration of electrons. A portion of the accelerated electrons is captured into a coronal loop system. The enhancement of intensity both the hard X-ray and EUV-emission in some points of another ribbon can be considered as the sites of electron precipitation (Fletcher and Hudson, 2001). 

In the context of superflares, we re-analyzed TRACE data in the EUV-ranges for this Bastille Day flare and found that the hard X-ray sources I, II, and III are footpoints of the corresponding coronal loops, e.g. they coincide with the EUV-sources. This coincidence of the sources is better seen just after the maximum of the hard X-ray burst. These TRACE data permit us to study an evolution of coronal loops linking source I with sources II and III. For example, Figure 2 shows a change of shape and height of a loop between sources I and III.  

Coincidences of the bright points in the ribbons in the EUV- and in hard X-ray ranges are seen in many powerful events, for example, on 28 October 2003, and even in the flare on 6 December 2006 located near the limb in not quite favourable conditions for observations (Krucker et al., 2011).

The idea by Livshits and Belov (2004) about efficient particle acceleration above spots was reliably confirmed by Grechnev et al. (2008) by an analysis of the whole set of observational data for the flare on 20 January 2005. In fact, many features of the largest solar flares, including motions, radiation in different spectral ranges and appearance of elementary particles are caused by this efficient acceleration of electrons and protons. 

	Before the discussion of a model of superflares, let us summarize briefly the knowledge on the appearance of the flare optical continuum. Short brightness enhancement was registered quite rarely in small regions of about a few arcsec. For instance, these white-light flares were seen distinctly in images, obtained at the Gyula Observing Station of the Debrecen Heliophysical observatory (with a filter of about 100\,\AA\ effective half-width centered at 5460\,\AA) (Kovacs, 1977; Dezs\"o et al. 1980; Babin, Dyatel and Livshits, 1985). These WL flares were located in close vicinity of spots or directly above them, and their brightness exceeded the brightness of the surrounding photosphere by up to 20--30\%. 
	
Note that most of the WL flares are characterized by fewer contrasts in the visible range. Zirin and Neidig (1981) proposed that this ``blue continuum'' including the Balmer jump and hydrogen lines originates from a thin layer of the hydrogen plasma above the chromosphere. Such a weak continuum is registered not only in bright patches near spots, but also in the $H_\alpha$-ribbons of powerful flares observed with TRACE. An analysis of the observations showed that the optical depth of WL flare source is close 0.01. This value was obtained by Potts et al. (2010) in the red spectral range, but it is typical also for the Yohkoh aspect camera in the blue range.

	The gas-dynamics modeling demonstrated that this radiation is associated with the downward moving low-temperature condensation and is the basis for interpretation of observations of line spectra of impulsive solar and stellar flares. The chromospheric response to an impulsive heating can be different depending on the energy and hardness of the electron beam. This explains a certain spread in properties of specific events, in particular, the appearance of the optical continuum in the condensation occurs rarely and only during the largest flares. It should be emphasized that the explosive evaporation occurs in a different way in small areas during the same flare (Sadykov et al., 2015). 
	
	However, the entire solar WL flare observations contain a few points that cannot be explained in the framework of the gas-dynamics model. First of all, this refers to classical observations about the high contrasts of WL flares in the red spectral range. An analysis of their spectral data requires adopting the optical continuum source that is situated in the upper photosphere. The opacity of layers beneath the temperature minimum is due to the negative hydrogen ion, $H^-$, and the hydrogen atoms. Any changes of physical conditions in these layers will be reflected in the intensity of red and IR-spectral ranges. 
	
	The red continuum began to be observed directly in spectra of powerful impulsive flares on red dwarf stars. So, Kowalski et al. (2013) presented observations of line and continuum emission from simultaneous high-cadence spectra covering near-ultraviolet and optical wavelengths (3400--9200\,\AA) and photometry for twenty M -dwarf flares. This analysis allowed them to trace the behaviour of the Balmer continuum at $\lambda < 3646\,$\AA\ and features of the line profiles both in weak and large flares. Observations in the blue spectral range agree well with both the two-temperature gas-dynamical modeling of an impulsive flare on the red dwarf (Katsova, Boiko, and Livshits, 1997) and the new single-temperature radiative hydrodynamic flare models by Allred et al. (2005). The most important result of the analysis of these observations by Kowalski et al. (2013) is the detection of two components of the flare optical continuum: blue continuum at the flare maximum that linearly decreases with wavelength from $\lambda=4000-4800\,$\AA, indicates the hot, black-body emission with typical temperatures of $T_{BB} \sim 9000 - 14\,000\,$K; the red continuum gives evidence for a $T_{BB} \sim 8000\,$K component at the beginning of the gradual decay phase during eight flares, most of which are impulsive events. These observations reveal some ``mysterious'' relationship between properties of the blue and red continua.  
	
The optical continuum, not related to the radiation of the low-temperature condensation, can appear on G-type stars at high energies of non-stationary processes apparently more often than during the large flares on the contemporary Sun. Note that the required heating of dense layers can be due to different causes. This may be associated with signatures of propagation of high-energy particles, in particular, electrons with $E> 500\,$keV, which can outstrip the main stream of the lower-energy particles, penetrate deeper and affect a larger area than the main flux of electrons with $E\approx 20\,$keV. In principle, the heating of the upper photosphere can be due to the return current (Matthews, Brown, and van Driel-Gesztelyi, 1998).

	So, what can be imagined about a model of the optical source of a superflare on G-type stars based on this experience of the study of solar phenomena? One clear result is the following: the continuum emission arises in loops with an area comparable with that of $H_\alpha$-ribbons of large solar flares. Most likely, three kinds of sources, I, II and III, similar to those presented in Figure 2, will stand out here. The emission of source III is due to gas-dynamical response to the particle flux, which fills up the coronal loop. Note that here particles can be both captured into an arcade and accelerated directly in the vertical current sheet as usually occurs in weak solar flares (Kopp-Pneuman model). The optical emission sources of ribbons are low-temperature chromosphere condensations in footpoints of flare coronal loops. 
	
In addition to this blue continuum and line emission, the optical continuum of the photospheric layers of the active region arises in low-lying loops near large spots where particles are accelerated (source I). This region close to the temperature minimum emits in a broad spectral interval from IR- to the visible range -- the red continuum with a low effective temperature. There are a few causes of small additional heating of the upper photosphere of G-type stars similar to those in red dwarfs. One of them can be associated with penetration of high-energetic particles from the acceleration site to the photosphere.
 
	An impact of the hardest particles on the temperature minimum region can also be substantial in source III. Therefore the ratio between the intensity of the blue and red continua can differ from one event to another. Besides, in some specific superflares, pairs I and III can occur and may be spaced in time. 
	
	It is possible to imagine the temporal development of a superflare. The start will most likely be a giant plasma ejection that either erupts from the stars as a CME or lifts up into the corona. At the beginning of the event, strong absorption arises due to the negative hydrogen ion, $H^-$, in source I, sometimes referred to as a black-light flare (van Driel-Gesztelyi et al. 1994; Ding et al. 2003). Then series of elementary episodes of particle acceleration develops accompanied by peaks of hard X-ray emission. Each of acceleration episodes provides the red continuum and the radiation of the condensation arises within a few seconds. As a result, the flare optical continuum extends from IR- to UV-ranges. 
	
Note that the total energy of the optical radiation of a superflare is the sum of the energies of elementary events, and each of them is characterized by the optical radiation in a broad spectral range. The radiation during the post-eruptive phase also contributes to the total flare energy of the superflare. Of course, the study of the photosphere response to an impulsive heating may enable us to concretize these predictions.

Returning to problems of superflares, note that their contrast on the stellar surface has to be close to that observed on the Sun. So, at the same total energy $10^{34}\,$erg, the observed flux ratio, $\Delta F/F = 10^{-3}$, corresponds to the same contrast as that in patches of WL flares on the Sun. However, on the Sun, these patches can be observed only rather closely to the particle acceleration site (source I). In superflares, the most energetic particles can reach another footpoint of the loop and penetrate there in deeper layers of the atmosphere before the main stream of particles with smaller energy would arrive there. This may enhance the low-temperature optical continuum.

Note that we do not discuss in full detail the radiative gas-dynamical simulations of processes in powerful impulsive flares. However, we have dealt with these problems earlier since 1980 (Livshits et al., 1981), and some years ago we carried out new gas-dynamics modeling based on the algorithm by Caunt and Korpi (2001). Using the same approach for radiative losses, we have obtained results close to previous our conclusions (Livshits et al., 1981). Advance in such a modeling for superflares is difficult because there is no final decision on the question how the return current limits the particle energy penetrating into the chromosphere. Some problems of an analysis of the non-stationary radiation field in a G star atmosphere where exists strong photospheric radiation requires amplification as well. 

However, before radiative gas-dynamical simulations of superflares, we analyzed the question of how the appearance of strong optical radiation can be explained in principle in the framework of the gas-dynamics model and what consequences follow from this consideration. Unlike most of other gas-dynamics models, we applied the single-fluid two-temperature approximation that allows us to get more exactly physical characteristics of a flare source in the first 1--2 seconds after the injection of the accelerated particle beam into the chromosphere and upper photosphere. This two-temperature model better describes the relationship between high-energy phenomena arising during the interaction of accelerated particles with the dense plasma. This model allows us tracing the connection between nuclear spallation reaction, microwave radiation, and subsequent formation of sources of the thermal X-ray and the optical emission. When the heating beam is hard, two sources of the optical continuum arise: the downward-moving chromospheric condensation, and a region of enhanced heating in the upper photosphere. Further we use the just one result of our gas-dynamics calculations that the flare continuum emission arises on the Sun (or the young G star) when the energy of the accelerated electrons is $3\times 10^{11}\,\hbox{erg}\:\hbox{cm}^{-2}\,\hbox{s}^{-1}$ and the differential spectrum $\gamma \le 3$. This minimal flux of the energy of electron provides the formation of the optical continuum that conventionally is indentified with the blue component found by Kowalski et al. (2013) for flares on red dwarfs. 

This chromospheric source is characterized by the maximal radiation area. But when accelerated electrons have a hard spectrum, the heating of the upper photosphere occurs. On the Sun, two types of white-light flares are considered: (i) hydrogen recombination continua, e.g. Balmer continuum similar to Figure 1 (see recent observations with IRIS by Heinzel and Klein, 2014), and (ii) photospheric white-light radiation due to photospheric temperature enhancement (e.g. backwarming), see review by Ding (2007). One can suppose that during superflares both sources of continuum emission arise. Their relative contribution to the total radiation varies from one superflare to another one.

It is important that this result allows us to evaluate the flux of accelerated electrons penetrating the chromosphere. The energy flux is fixed while the spectrum of electrons with $E > 20\,$keV can have different power law indices. However, in this case the mean energy of the beam is around 23--26\,keV, and for the maximal electron flux $3\times 10^{11}\,\hbox{erg}\:\hbox{cm}^{-2}\:\hbox{s}^{-1}$, we obtain the electron flux $7.5\times 10^{18}\,\hbox{electron}\:\hbox{cm}^{-2}\:\hbox{s}^{-1}$ and the total number of electrons that enter the chromosphere is $7.5\times 10^{37}\,\hbox{electron}\:\hbox{s}^{-1}$ in this superflare. Apparently, injection of accelerated electrons with $E > 20\,$keV into the coronal part of the source will have the same order of magnitude. For M5--X1 solar flares, this value is smaller by a factor $\sim 100$.

This is a parameter determined by the X-ray and radio emission of flares. Here of importance are the hardness of this electron beam, the magnetic fields, the temperature and the density of the plasma in the radiation sources.

In the case of the hard spectrum of accelerated electrons, this model has several consequences, which can be verified in current observations. This point will be discussed in the next two sections.

\section{The microwave radiation of flares}

The optical observations of superflares with $E = 10^{34}\,$erg give information about the total number of electrons with $E > 20\,$keV ~ $N\approx 10^{38}\,\hbox{electron}\:\hbox{s}^{-1}$. It is clear that this energy should be provided by primary injection of accelerated electrons from the acceleration site into the coronal loop. If the electron spectrum is soft, then gas-dynamical processes will heat the corona.

\begin{figure}[!t]\centering                      
\includegraphics[height=0.7\textheight]{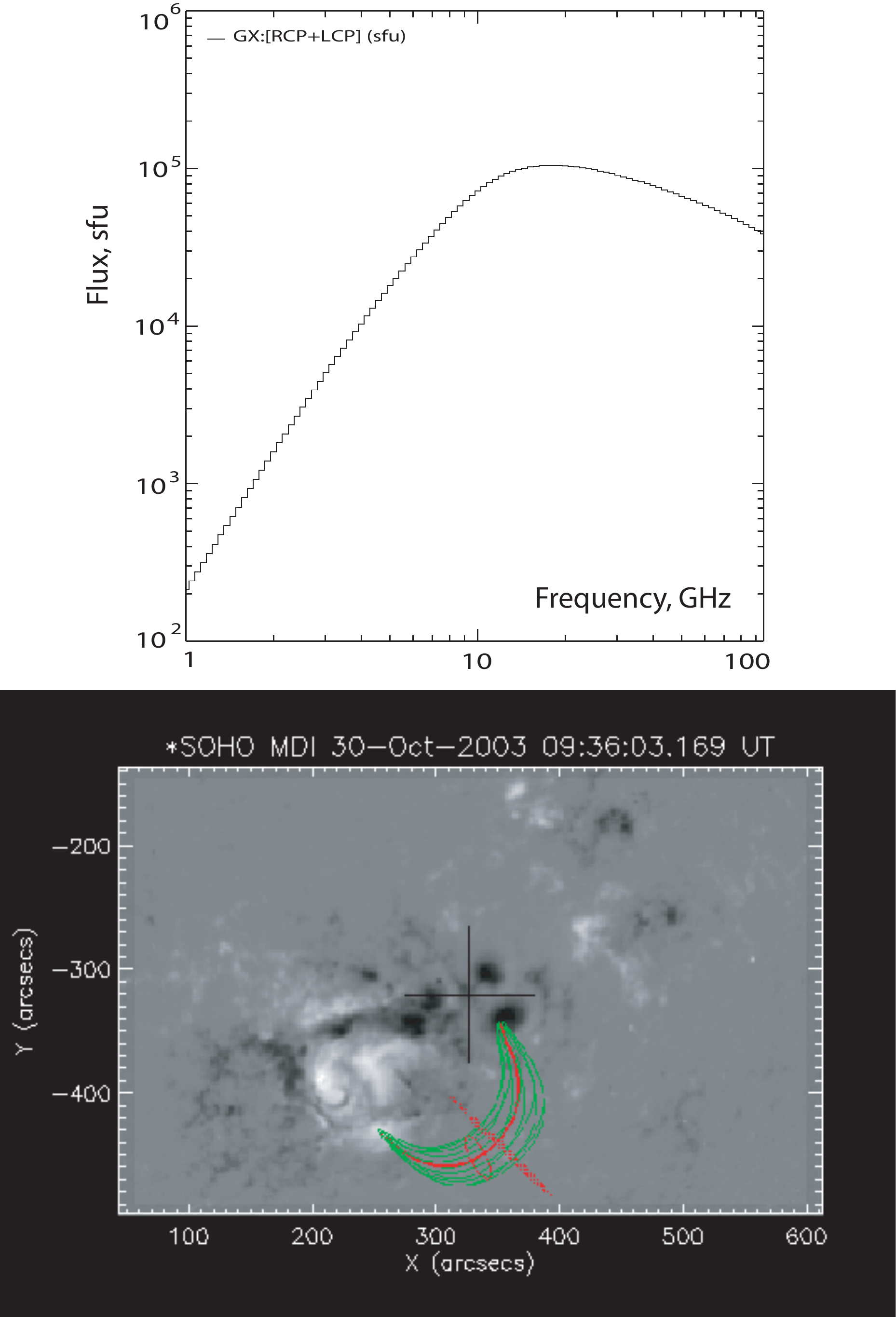}
\caption{\\
The calculated microwave spectrum for the solar flare on 28 October 2003 (upper panel). The calculation parameters are given in the text. The lower panel presents the location of the loop in the MDI image.
}
\end{figure}

The soft X-ray radiation which arises during such an event is high enough to be observable at stellar distances. Even if the electron spectrum is hard, the X-ray radiation at the range above 20\,keV from stellar flares is not yet available for observations. But in the case of the hard spectrum, the same total number of accelerated electrons will result in effects that can be registered and be a test to verify our assumption that the powerful optical radiation of superflares is associated with effective particle acceleration.

We evaluate further a level of the microwave radiation of superflares. First, the gyrosynchrotron radiation is well-studied in large solar flares. The spectrum of the microwave burst has a maximum at frequencies of around 10\,GHz, but in solar flares larger than M3 (with possible particle acceleration), this maximum is shifted toward the higher frequencies. The maximal fluxes at 35 GHz are within a wide range $I_{35} = 1-100\,000\,$sfu (1 sfu$=10^{-22}\,\hbox{W}\:\hbox{m}^{-2}\:\hbox{Hz}^{-1}$). The IZMIRAN data base for observations in 1976--2006 with different radio telescopes shows that this maximal flux $I_{35}$ at this frequency weakly depends on the soft X-ray flux, $I_R$, for faint flares (where the thermal radiation is still significant), while the $I_{35}$ value is proportional to $I_R^{1.5}$ for flares with particle acceleration. 

	We are interested here in the greatest solar flares. The high-energy phenomena in some of them were studied in detail in the past solar cycle. It is clear that microwave bursts arise if electron beams contain a large amount of microwave-emitting electrons with $E > 0.5\,$MeV. The largest events with $E \ge 10^{32}\,\hbox{erg}$ are accompanied by 3--20\,GHz bursts with a peak flux up to $10^5\,$sfu. A similar flux was registered for the flare on 20 January 2005, where footpoints were situated above the sunspots of the opposite polarity (Grechnev et al., 2008). Because this big flare occurred at 60\,W, this case is not best suited for extrapolating the magnetic field into the corona. We have selected another large flare on 28 October 2003 near the disk center. Unfortunately, RHESSI was disabled right during the maximum of this flare $>$X17, but its X-ray radiation was registered by SONG/KORONAS-F, HEND/Mars-Odyssey and some radio telescopes. Using the code for computing the gyrosynchrotron radiation (Fleishman and Kuznetsov, 2014; Nita et al., 2014), we calculated the spectrum of the microwave radiation of the accelerated electrons with the single slope $\gamma= -2.5$. The magnetic field in the corona was reconstructed from a SOHO/MDI magnetogram. The microwave emission arose in the loop of length of $2\times 10^{10}\,$cm and width of $3\times 10^9\,$cm. The magnetic field strength in the loop top was 175\,G. In this case it is impossible to achieve consent between observed and calculated spectrum (the flare on 20 January 2005 is shown in Figure 11 in Grechnev et al., 2008, and our modeling by another code also confirms this).
	
In spite of similar shapes of microwave spectra of both flares, the soft X-ray radiation of the flare on 28 October 2003 exceeded the corresponding value for the flare on 20 January 2005, and the maximal microwave fluxes of the second flare were somewhat higher, by a factor of 1.5--2. Therefore, we report the version of computations of the emission in the reconstructed shape of the flare loop for the 28 October event, but with corrected values of the accelerated electron flux. The result is presented in Figure 3. The calculation is carried out for the following parameters: the density of electrons with $E > 20\,$keV is $3\times 10^7\,\hbox{cm}^{-3}$, the density of the thermal electrons is $n_0 = 5\times 10^9\,\hbox{cm}^{-3}$ in the loop and $1\times 10^9\,\hbox{cm}^{-3}$ in the surrounding corona. The density of thermal electrons decreases according to the exponential law for $T = 2\times 10^7\,$K and $1\times 10^6\,$K in the loop and surrounding corona correspondingly. Note that the total number of accelerated electrons in a coronal loop is $3.5\times 10^{35}$. We believe that these conditions describe the situation in the greatest hard flares on the Sun when the maximal microwave flux reaches $10^5$\,sfu at 10--100\,GHz. 

Now we can evaluate the microwave flux in superflares on G-type main sequence stars. Here certain information about solar phenomena, namely, the statistical proportional dependence of $I_{35}$ on $I_R^{1.5}$ for flares with particle acceleration can be helpful. The $I_R$ value for flares is closely related to their total energy. If we adopt the total energy of the largest solar flares to be $10^{32}\,$erg and use the same law for variations of the microwave emission in the form of $I_{35}$ is proportional to the total flare energy $E_{tot}^{1.5}$, then we obtain that the microwave flux of flares with $E = 10^{34}\,$erg at the distance of 1\,AU is about $10^8\,$sfu. This means that such a flare on a star at a distance of 100\,pc will produce a microwave flux above 2\,mJy.

         So, a microwave flux of 80\,mJy was registered at the peak of the optical flare on the young star V773\,Tau with NAOJ Nobeyama at 98, 64, 43, 23\,GHz (Iizuka et al. 2014). However, this phenomenon relates rather to non-solar analogues. It would be of importance to carry out the monitoring with VLA and ALMA of some of G-type stars with a high occurrence of superflares.

\section{Lithium production during impulsive flares}

Let us consider again phenomena associated with the efficient particle acceleration, when the large number of not only electrons but also protons can be accelerated up to high energies, in particular, protons can reach energies of up to dozens of MeV. The high-energy protons interact with nuclei of the ambient photosphere and a series of spallation reactions like the $\alpha-\alpha$ process is developed. A review of the nuclear processes in solar flares was presented by Ramaty (1986). From observation of the $^7$Li line at 0.478\,MeV, Livshits (1997) gave an estimate of the amount of lithium produced in a large impulsive flare on the Sun. He concluded that about $10^{30}$ Li nuclei can be produced in the largest impulsive flares on the Sun. It was also shown that the diffusion of new Li ions is slow under the conditions of the turbulent convection and the produced excess of the Li atoms can be observable during several hours after the flare. Hereafter Ramaty et al. (2000) found the nuclear reaction where accelerated $^3$He reacts with $^4$He, which can lead to production of the $^6$Li nuclei during flares. 

There are no direct observational evidences for this effect in solar flares. But Montes and Ramsey (1998) reliably found the enhancement of the resonant Li 6708\,\AA\ line during the long-duration optical stellar flare. The Li line equivalent width followed the temporal $H_\alpha$-profile of this flare. The idea of the production of Li by spallation reactions is confirmed by the observations of the rise of the $^6$Li/$^7$Li ratio in this flare. Note that this flare occurred on an active K0 star with a rotational period of 9.98 days and lasted for 3 days. These data were reviewed by Cutispoto (2002) in the context of the interrelation of lithium and other light elements and stellar activity. It is argued that one should expect that the $^6$Li and $^7$Li isotopes arise during large flares, but the abundance of $^7$Li nuclei are not seen to change against the abundance of Li atoms originating from the stellar interiors, where the proton-proton reactions occur, while the $^6$Li atoms arising in the flare are manifested and lead to a change in the $^6$Li/$^7$Li ratio. 

A new approach to the study of the question whether new Li has been produced by nuclear reactions during flares was proposed by Flores Soriano, Strassmeier, and Weber (2015). They observed in spectra of the young active K2 dwarf LQ Hya (with a rotation period $P_{rot}=1.6\,$d) ``that the strength of the Li line and other temperature sensitive lines is rotationally modulated and clearly enhanced in the hemisphere of the star with the highest magnetic activity'', i.e., on the hemisphere with the larger spot area. Generally speaking, this modulation is associated with an increase in average temperature. But on the same hemisphere, the observed spectral features are interpreted by Flores Soriano, Strassmeier, and Weber (2015) as the appearance of new $^6$Li nuclei. Because ``sporadic flares of moderate intensity were also detected'' at the same hemisphere in the hydrogen and helium lines, the authors supposed that these Li atoms are produced during flares.
 
One can expect a significant amount of lithium being produced during superflares. The set of spectral observations of the equivalent width of the resonant Li 6708\,\AA\ line, $W_{Li}$, can help to reveal a certain part of temporal variations with the rotation period associated with starspots. The detection of another, irregular variability of $W_{Li}$ will indicate a flare. Both the duration and shape of the temporal profile of the equivalent width variations differ significantly. Such observations require long-term spectral monitoring with large telescopes.

\section{On the evolution of activity of late-type stars}

Where can superflares with the total energy $E > 10^{34}\,$erg occur? To address this question, we discuss briefly some results of the study of activity of low-mass G0--K5 stars. 
As it is known, a protostar spins faster during a stage of the gravitational contraction, and the primary relict magnetic field is enhanced. Then the rotation of a star is decelerated, and some activity processes are developed, as it follows from observations of star -- members of open clusters of ages of 30--600\,Myr. 

The dependence of the X-ray radiation of late-type main-sequence stars on the rotation period shows that activity of fast rotating stars is saturated, the coronal activity index, $R_X = \log L_X/L_{\it bol}$ is close to --3 and weakly depends on the rotation period (Pallavicini et al., 1981; Wright et al., 2011). Stars with moderate activity up to the solar level demonstrate the clear relationship $\log L_X \sim P^{-2}_{\it rot} \sim \nu^2$, where v is the rotation rate (Pallavicini et al., 1981; Reiners, Sch\"ussler, and Passegger, 2014). The rotation rate is a direct indicator of both stellar age and activity (Skumanich, 1972). The further analysis of the ``rotation--age'' and ``activity--rotation'' relationships allowed Mamajek and Hillenbrand (2008) to propose a method of gyrochronology for estimate of a star's age from its activity level. The transition from young stars with activity saturation to less active, older stars happens when their rotation slows down to periods of 10 days. Periods of 7--12 days are characteristic both for the first types of stars and for another group of intermediate age stars.
 
In order to investigate the activity of stars older than 600 Myr, we considered data on chromospheric activity, namely, the chromospheric activity index, $\log R'_{HK}$, that is the ratio of the emission components of the H and K Ca II line to the optical continuum at 4000\,\AA. We analyzed the results of the HK Project of the monitoring of long-term chromospheric variations of more than 100 late-type stars in the solar vicinity since 1965 (Baliunas et al., 1995). Besides, we added these data to observations obtained during the Planet Search Program (Wright et al., 2004; Arriagada, 2011) and built a ``chromosphere--corona'' diagram for the chromospheric, $\log R'_{HK}$, and coronal activity indices $\log L_X/L_{\it bol}$ (Figure 4). Then we included observations of stars with high proper motions (Maldonado et al. 2010) and stars with detectable lithium abundance (Mishenina et al., 2012). The location of the young stars with activity saturation is indicated by an oval according to the data of Martinez-Arnaiz et al. (2011).

\begin{figure*}[!t]\centering                      
\includegraphics[width=\linewidth]{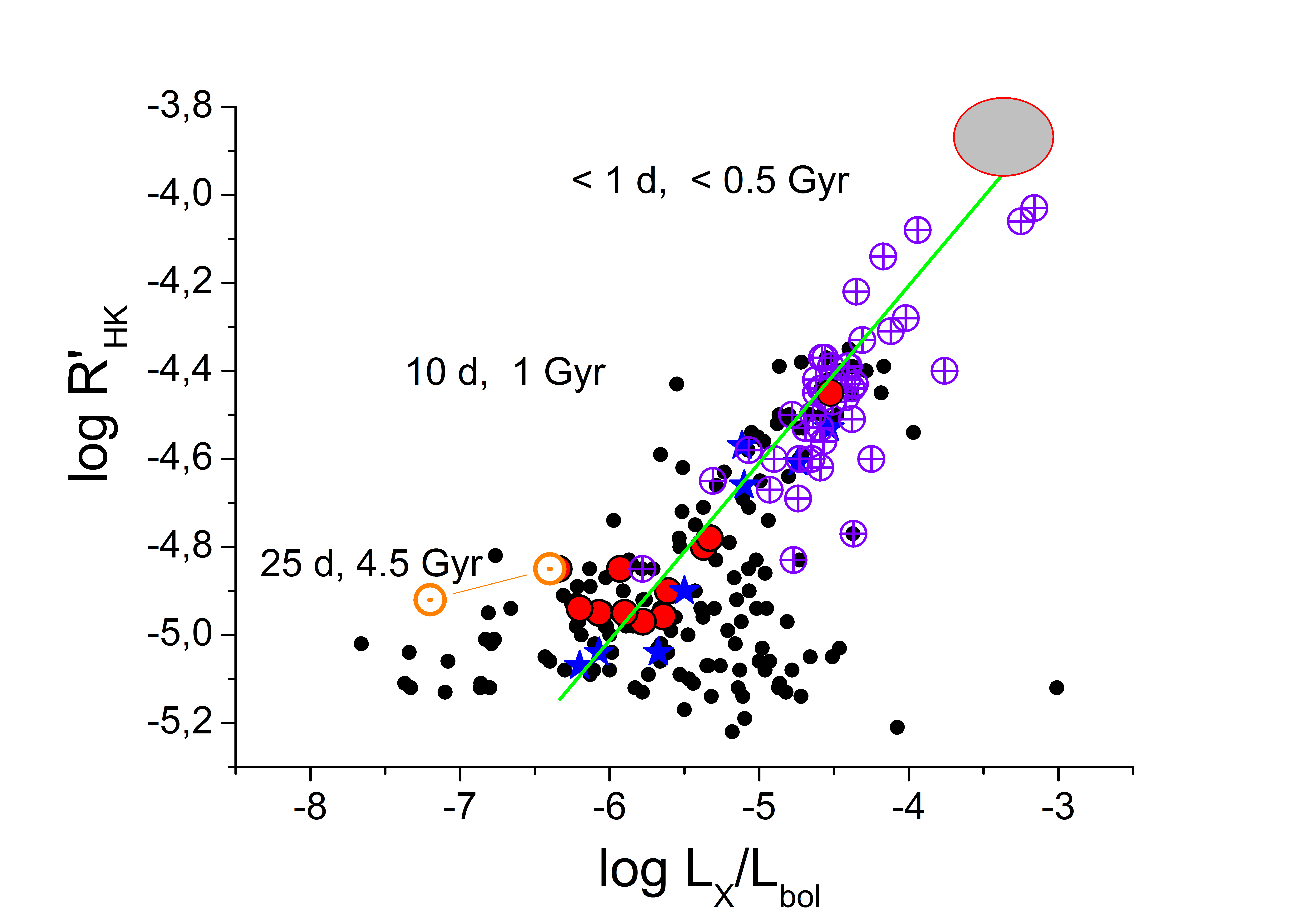}
\caption{\\
The ``chromosphere--corona'' diagram of the chromospheric and coronal activity indices for late-type stars. Stars with detected lithium abundances are marked as violet crosses inside the circles (similar to Figure 9 in Mishenina et al., 2012). The stars of the basic data set are marked by black dots. According to a kind of the cycle, the stars with \textit{``Excellent''} cycles are marked as red circles, the stars with \textit{``Good''} cycles are indicated as blue asterisks; the Sun at maximum and minimum is denoted by its own sign, connected by a straight line. The location of stars with saturated activity is conditionally designated with oval. The straight line corresponds to a single-parametric gyrochronology.
}
\end{figure*}

At present it appears that dating of star's ages from data of open clusters is more reliable (Meibom et al., 2015, and references therein). This analysis allowed us to improve the ``chromosphere--corona'' diagram. Figure 4 shows that the basic group of stars is concentrated near the straight line in the diagram that connects young stars of ages of hundred million years to the older stars whose age is comparable to that of the Sun (4.5 Gyr) or more. It is clearly seen that the distribution of young stars with activity saturation finds its continuation in the distribution of the stars with high lithium abundance. Strongly spotted stars, most of them are the BY Dra-type stars, are situated in a compact group on the ``chromosphere--corona'' diagram around the values $\log R'_{HK} = -4.5$ and $R_X = -4.5$. The stars, whose activity most closely matches what is observed in the contemporary Sun, are located below this point, along the straight line. There is reason to believe that activity of all stars from this level to activity of the contemporary Sun is caused by local magnetic fields with a definite governing role of the large-scale magnetic field. In particular, it relates to such known active stars as HD 1835 (BE Cet), HD 20630 ($\kappa^1$ Cet) and HD 152391 (V 2292 Oph). The relative spot area of the G star, BE Cet, reaches 3.3\% (Alekseev, 2001). The G7 star, V 2292 Oph, is characterized by the faster rotation than for other stars with \textit{Excellent (well-detected)} cycles. In recent years, it has been noted that the cyclic activity with a period of about 11 years became less regular in this star. This allows us to consider it as an example of a star with a \textit{Good (established)} cycle. (The classification of stellar cycle types was introduced in the HK Project by Baliunas et al., 1985). The rotation periods of these stars are 8, 9, and 11 days, respectively. There is reason to consider that the age of main-sequence stars rotating with periods of about 10 days is slightly older than 1 Gyr. We call such a main-sequence G star ``a young Sun'', and for active processes on stars of ages from young suns to the contemporary Sun, use the term ``solar-type activity''. Note that this term involves the existence of a cycle on a given star. In fact, the diagram shows two basic groups of stars with activity close to the saturation level and to the solar-type one (sees Figure 4 in Reiners, Sch\"ussler, and Passegger, 2014, where these groups are shown separately). These problems of the evolution of activity were discussed in more details by Katsova (2012). Note that activity of stars situated below the straight line, that describes single-parametric gyrochronology, can evolve through another path (Katsova and Livshits, 2011). 

On the stars with activity saturation and stars where Li is reliably detected, the rotational modulation of the chromospheric and coronal emission is clearly pronounced. This is due to the concentration of active regions to certain fixed longitudes. Unlike for stars with solar-type activity, large-scale magnetic fields have strong effects here.

The clearest result of the analysis of the Kepler superflares is the fact that these phenomena mainly occur on fast-rotating stars. For instance, if we plot the number of superflares versus the rotational periods of stars, then one can see the main maximum at periods of 1--5 days and the second one at 12--13 days. The corresponding histogram showing the minimum of the flare number is given by Katsova and Livshits (2014, Figure 5 therein). Thus, the superflares occur on G stars of the first group with activity saturation or close to this level. Only a small fraction of superflares occur on G stars with solar-type activity and a strong spottedness.

\section{Magnetic fields of G stars and  maximal flare energies}

Until now, variations in the magnetic field strength with the axial rotation phase together with spectral data provide an opportunity to simulate the distribution of inhomogeneities over the surface of several active G and K main-sequence stars. This method of Zeeman Doppler mapping allows us in the first approximation to separate the contributions of spots and of the large-scale magnetic field. 

New data on the magnetic fields of F, G, and K stars were recently obtained in the frameworks of the \textit{``Bcool Collaboration"} (Marsden et al., 2014). The spectropolarimetric observations of 170 late-type stars were carried out with large telescopes. The mean surface longitudinal magnetic field, $B_l$, was detected in about 40\% of the \textit{``Bcool''}-sample stars. This value represents the $B_l$ signal averaged over all observations of the star. 

We analyzed these observations for G stars of different ages only. We chose all G stars with values $B_l$ exceeding $3\sigma$ from Table 3 of Marsden et al. (2014). From that list, we excluded some of the rapidly rotating and hence very young stars with strong magnetic fields. Thus, the final list includes 28 G stars whose rotation periods are $>5\,$days. The result is presented in Figure 2 in Katsova and Livshits (2014). We obtained the average value of the modulus of the field $|B_l| = 4.72 \pm 0.53\,$G for G stars. The obtained average value corresponds to a rotation rate of about 4 km s$^{-1}$, i.e., twice faster than the rotation of the modern Sun. This allows us to adopt such a value of the field as an average value for a G star with a rotation period of 10 days, which corresponds to the age of about 1 Gyr. 

We compared this value with data on the general magnetic field of the Sun as a star. Values averaged over the Carrington rotation near the maximum of the cycle do not exceed 0.5 G (see, for example, Kotov, Haneychuk, and Tsap, 1999 for data on the magnetic field of the Sun as a star in 1991). Thus, the average magnetic field strength of young G stars is at least higher by an order of magnitude than that for the maximal Sun in the modern era.

Note that this evaluation is supported by the fact that the magnitude of $|B_l|$ for stars whose parameters are most similar to those of the young Sun, is 8.5 G for V 2292 Oph and 7.7 G for $\kappa^1\,$Cet (other data for this star  $|B_l|=7.0$, do Nascimento et al., 2014). 

Now there is certain information on the structure of the large-scale magnetic fields of late-type stars. In particular, it was found that slowly rotating stars have magnetic fields with the same structure as that on the Sun, which are regularly changing during the 22-year magnetic cycle. In other words, the global dipole does exist virtually all the time, and only during the polarity reversal shifts the magnetic equator in the latitude, which can be considered as a manifestation of a large-scale toroidal component. Active regions are located rather chaotically, active longitudes are allocated poorly, and toroidal fields are characterized by comparatively small scales (local fields). At the same time, the large-scale toroidal component is clearly pronounced on G stars with rotation periods shorter than 12 days (Petit et al., 2008). Although it prevails over large-scale poloidal magnetic fields, the difference of the magnetic field strength is small. So far it is difficult to relate these conclusions with new data on spots on the rapidly rotating G stars.

Thus, we conclude that the general nature of the magnetic fields on G and K stars is similar to that observed on the Sun now. Namely, there is a large-scale magnetic field, most clearly manifested in the polar regions and in the local fields at low latitudes. The magnetic field strength of 1--3 kG was found in spots covering up to 10\% of the stellar surface (see Saar, 1996, Gershberg, 2005). 

 The development of non-stationary processes in solar active regions is associated not with the field energy of a given magnetic configuration, but with its distinction from the potential energy. It is characterized by the free energy. It increases with an emergence of a new magnetic flux and it is probably spent in flares and CMEs. It was found that on the Sun the free energy is accumulated at altitudes of about 2000 to 5000 km (Sun et al., 2012) and 10--15\% of the free energy of a given AR can be spent on a large flare in it. For four large flares this follows from results by Jing et al. (2009), and for dozens of other flares, this result can be derived from Falconer et al. (2011). Both these conclusions are based on an extrapolation of the photospheric magnetic field into the corona (Metcalf et al. 2008, where authors applied nonlinear force-free field extrapolation algorithms) and agree with our calculation in Livshits et al. (2015) as well. 
 
A non-stationary process starting immediately above the dense chromosphere layers quickly extends to the coronal part of an AR. Accelerated particles fill up one or a few coronal loops. Basic sources of the hard X-ray radiation are located in footpoints of the loops. Note that a detailed discussion of the location of hard X-ray sources, ribbons in the chromosphere and white-light emission is given by Krucker et al. (2011), and one can only agree with their statement that ``future observations should focus on the three dimensional structure of the sources''.

The measured average magnetic field of a star is composed of the large-scale field and the field of spots. Like the Sun, the field of individual spots reaches a strength of several kG, but spottedness of the stellar surface, filling-factor, in rapidly rotating stars exceeds that of the Sun by 1--3 orders of magnitude. Finally, the contribution of small-scale and large-scale fields to the measured average magnetic field of a star is about the same (Petit et al., 2008; Marsden et al, 2014). According to measurements of the fields on individual stars, and on a large group of stars, we can conclude that the magnetic fields on G-type dwarfs have approximately the same general character as that of the Sun, but the strength of the field in the vicinity of neutral lines, separating the oppositely directed magnetic field, is one order of magnitude larger than that on the Sun.

This evaluation corresponds to an increase in the field energy above the AR on G-type dwarfs by two orders of magnitude as compared to large ARs on the Sun. Repeating the arguments similar to those used above for the Sun, we obtain an upper limit of the total flare energy on young Sun (with an axial rotation period of 10 days, that corresponds to an age of about 1 Gyr) equal to $10^{34}\,$erg. This evaluation is obtained by application of the magnetic virial theorem (Livshits et al., 2015). 

As for flares of higher energies, from $10^{34}\,$erg to $10^{36}\,$erg, there is evidence for occurrence of these events on stars with a radius larger than one solar radius. This correlation has been revealed by Balona (2015). If the radius of the G star increases, then the characteristics of activity, including flares, changes significantly. 

Of course, superflares can have a nature that differs from non-stationary processes on the Sun. In particular, they may be a result of the interaction between planets and the host star: these problems were discussed in Rubenstein and Schaefer (2000); Cuntz, Saar, and Musielak (2000) and Ip et al. (2004).

\section{Concluding remarks}

Before the detection of superflares on stars, only one event was registered on a red dwarf that differed from the solar analogues and indicated a possible realization of nuclear spallation reactions. So, the simultaneous X-ray and optical observations of a giant flare on the red dwarf CN Leo by Schmitt et al. (2008) and Fuhrmeister et al. (2008) demonstrated the definite difference of this phenomenon from typical impulsive flares on red dwarfs and on the Sun. In particular, instead of motion of the low-temperature condensation downward to the photosphere, an explosion of the coronal and chromospheric plasma was registered in the first seconds of the flare onset. 

In the context of finding super powerful phenomena on G stars, we try to explain where and how such events can occur. These observations were carried out in the optical. Until now, an interpretation of these superflares was associated with direct solar analogues (Shibata et al., 2014). In light of the fact that these phenomena are caused by particle acceleration, we provide observational evidence for the study of the origin of these superflares. We suppose that the optical continuum mainly arises during the explosive evaporation of the chromosphere and estimate an area of the optical flare as high as $10^{19}\,\hbox{cm}^2$ for a total flare energy of $10^{34}\,$ergs. This means that all flare ribbons emit in the optical continuum differentlty from what on the Sun. This radiation is enhanced in regions above the $H_\alpha$-ribbons where particles are accelerated, and also in the second footpoint of the main loop. 

If this electron flux is hard (slope $\gamma$ is 2--3), then the microwave emission can reach 2 mJy at a distance of 100 pc, which can be detected by modern radio facilities. Secondly, protons above 30 MeV and $\alpha$-particles when interacting with ambient particles produce light elements, in particular, lithium due to a series of spallation reactions (Ramaty, 1986; Livshits, 1997; Ramaty et al., 2000). There is great interest in the detection of an enhancement of the resonant Li I line in superflares. 

	One of the important results of the Kepler observations is a statistics of the flare occurrence frequency on G stars: superflares are observed more often on the faster rotating stars. We have analyzed in more details which stars demonstrate these powerful phenomena from the point of view of the evolution of solar-type activity. Among the active late-type stars there are young objects with saturated activity, which weakly depends on the rotation rate. There is another group of main-sequence stars where the chromospheric and coronal activity is determined by the axial rotation. Many arguments indicate that phenomena on stars with a rotation period $>10\,$days can be considered as solar-type activity. In particular, here the interconnection between large-scale and local magnetic fields changes, while their role in the formation of the activity varies as well. Note that the young Sun with a rotation period of about 10 days is about 1 Gyr old. Namely, the solar-type activity has already formed by this age. 
	
	As is known, on the Sun the free energy of the magnetic field accumulates in the chromosphere and is then released in non-stationary processes. Powerful events with effective particle acceleration occur in regions where the polarity inversion line (PIL) passes sites with a strong magnetic field (near spots). It is clear that the total flare energy in a given active region cannot exceed the free energy therein. Because the free energy is released in several steps, every large flare ``devours'' 10--15\% of the deposited free energy in this region. These estimates derived from data on the full vector of the magnetic field agree well with the energies of largest solar flares, $10^{32}\,$ergs.
	
	 We have analyzed new observations of magnetic fields on G stars with the rotation periods $> 7\,$days. The mean longitudinal magnetic field is 10 times stronger than that on the Sun at the maximum of the cycle. This allows us to estimate that the maximal energy of flares on these G stars is not more than $10^{34}\,$ergs. 
	 
In conclusion, we would like to emphasize that the interpretation of the observations of superflares on G stars provides a promoted understanding of extreme processes on the Sun. Beside, the detection of the flare microwave source and the emission in the Li resonant line could demonstrate how effective particle acceleration can be on stars in the lower part of the main sequence.

\phantomsection
\section*{Acknowledgments}

We are grateful to the Organizing Committee for the possibility of presentation of our results on "Solar and Stellar Flares" at the conference in Prague, June 2014. We are thankful to V. V. Grechnev, T. I. Kaltman for help in an analysis of microwave data, and to A.B. Belov and B.A. Nizamov for discussions. We are deeply grateful to the referee for the helpful remarks.

This work is supported by Russian Foundation of Basic Research under grants No 14-02-00922, 15-02-06271 and by the grant from the Russian Federation President supporting Scientific Schools 1675.2014.2.

\phantomsection
\bibliographystyle{unsrt}

\end{document}